\documentclass[11pt]{article}
\usepackage{graphicx}
\renewcommand{\theequation}{\arabic{section}.\arabic{equation}}
\def\lb{\label}
\def\bb{\bibitem}
\def\be{\begin{equation}}
\def\ee{\end{equation}}
\def\ba{\begin{eqnarray}}
\def\ea{\end{eqnarray}}
\def\ol{\overline}
\def\e{{\rm e}}
\def\R{{\cal R}}
\def\nn{\nonumber}

\begin{document}
\begin{titlepage}
\title{
A scenario for critical scalar field collapse in $AdS_3$}
\author
{G\'erard Cl\'ement\thanks{Email:gerard.clement@lapth.cnrs.fr}\\
\small{LAPTh, Universit\'e de Savoie, CNRS, 9 chemin de Bellevue,} \\
\small{BP 110, F-74941 Annecy-le-Vieux cedex, France} \\ Alessandro
Fabbri\thanks{Email:afabbri@ific.uv.es}\\ \small{Centro Studi e
Ricerche Enrico Fermiâ Piazza del Viminale 1, 00184 Roma, Italy}
\\ \small{Dipartimento di Fisica dell'Universit\`a di Bologna,
Via Irnerio 46, 40126 Bologna, Italy}\\ \small{Dep. de F\'isica
Te\'orica and IFIC, Universidad de Valencia-CSIC,}\\ \small{C. Dr.
Moliner 50, 46100 Burjassot, Spain}}
\date{}
\maketitle

\abstract{We present a family of exact solutions, depending on two
parameters $\alpha$ and $b$ (related to the scalar field strength),
to the three-dimensional Einstein-scalar field equations with
negative cosmological constant $\Lambda$. For $b=0$ these solutions
reduce to the static BTZ family of vacuum solutions, with mass $M =
-\alpha$. For $b\neq0$, the solutions become dynamical and develop a
strong spacelike central singularity. The $\alpha<0$ solutions are
black-hole like, with a global structure topologically similar to
that of the BTZ black holes, and a finite effective mass. We show
that the near-singularity behavior of the solutions with $\alpha>0$
agrees qualitatively with that observed in numerical simulations of
subcritical collapse, including the independence of the
near-critical regime on the angle deficit of the spacetime. We
analyze in the $\Lambda=0$ approximation the linear perturbations of
the self-similar threshold solution, $\alpha=0$, and find that it
has only one unstable growing mode, which qualifies it as a
candidate critical solution for scalar field collapse.}

\end{titlepage}\setcounter{page}{2}

\section{Introduction}
In 1993 Choptuik \cite{chop} studied the spherically symmetric
collapse of a massless scalar field in four-dimensional Einstein
gravity and found numerical evidence of a critical behaviour at the
threshold of black hole formation. The black hole threshold is
reached when a certain parameter $p$ characterizing the initial data
reaches a critical value $p_*$. The critical behavior, since then
observed in other sytems and various spacetime dimensions (for a
review, see \cite{gund}), is characterized by the emergence of a
(continuous or discrete) self-similarity and a power-law scaling of
the black hole mass $M\sim (p-p_*)^{s\gamma}$ where $\gamma$ is a
universal exponent, and $s$ depends on the dimension. Inspired by
this work, several years later Pretorius and Choptuik \cite{PC} and
Husain and Olivier \cite{HO} performed numerical simulations of
circularly symmetric gravitational collapse of a massless scalar
field minimally coupled to $2+1$ dimensional Einstein gravity with a
negative cosmological constant $\Lambda$. These simulations showed
the appearance of a critical regime near the theshold of black hole
formation, with continuous self-similarity and power-law scaling. A
class of exact continuously self-similar solutions of the
$\Lambda=0$ equations, depending on an integer $n$, was constructed
by Garfinkle \cite{G}, who found that the $n=4$ solution was a good
fit to the numerical data of \cite{PC} for critical collapse near
the singularity (where the effect of the cosmological constant is
negligible). The linear perturbation analysis of the Garfinkle
family of solutions, performed in \cite{GG}, showed that with
suitable boundary conditions the $n=2$ Garfinkle solution admitted a
single growing mode, suggesting that this solution should be the
critical one near the singularity. In \cite{CCF}, the Garfinkle
solutions were extended to solutions of the full field equations
truncated to first order in the cosmological constant $\Lambda$, and
the zeroth order linear perturbation analysis of \cite{GG} was
extended to the same order, confirming the results of that analysis.

However, several points remained obscure. Is there really a good
reason to prefer the $n=2$ solution over the $n=4$ one, which fits
better the observations \cite{GG,gund}? Also, the knowledge of the
critical solution in the zeroth order, or even in the first order,
approximation sheds no light on the global structure of the exact
critical solution. What we feel is perhaps a more important drawback
of the Garfinkle approach is its inability to explain a peculiar
feature uncovered in the analysis of \cite{PC}. They observed that
the introduction of a point particle, in other words a central
conical singularity, left unchanged (up to a phase shift in proper
time) the critical solution, even for an angle deficit very close to
$2\pi$. This suggests an alternative scenario in which the critical
solution, instead of having a regular timelike center as assumed in
\cite{G}, should more probably have no timelike center at all, in
close analogy with the case of the BTZ vacuum (or extreme) solution,
which lies at the threshold between BTZ black holes with positive
mass $M$ and no center, and AdS spacetime, generically with a
conical singularity (unless $M=-1$).

Recently, an analytical solution to the Einstein-scalar field
equations with $\Lambda<0$ was derived from a self-similar ansatz,
and argued to be relevant to critical collapse \cite{BST}. However
this argument was shown in \cite{comcrit} to be incorrect as
presented. It is nevertheless suggestive, in view of the above
remarks, that this solution is conformal to the three-dimensional
Minkowski cylinder (Minkowski space with one of the cartesian
spatial coordinates periodically identified), and thus has no
center. The purpose of the present paper is to further explore the
properties of this solution and those of a class of neighboring
exact solutions. We shall show that, while a number of these
properties agree qualitatively with those expected for the critical
solution, the quantitative comparison with the numerical simulations
shows that it cannot be the correct critical solution. The
shortcomings of this solution are probably due to its lack of an
asymptotic AdS region, presumably a direct consequence of its being
a separable solution. Nevertheless, we believe that our analysis
points the way towards a possible exact critical solution within the
no-center scenario for scalar field collapse in $AdS_3$.

In Sect.\ 2 of this paper, after briefly recalling the separable
solutions constructed in \cite{crit2}, we focus on a class of
solutions depending on two parameters $b$ (the scalar field
strength) and $\alpha$ real. We first show that the vacuum solutions
($b=0$) of this class correspond respectively to sectors of the BTZ
black hole spacetime with mass $M=-\alpha$ for $\alpha<0$, of the
BTZ vacuum for $\alpha=0$, and of the AdS spacetime with a conical
singularity for $\alpha>0$. The solutions with $b^2\neq0$ have in
common a strong spacelike central singularity, but differ by their
global structure. We show that the $\alpha=0$ solution of \cite{BST}
lies at the threshold between dynamical black-hole like solutions
for $\alpha<0$ and solutions with a naked timelike central conical
singularity for $\alpha>0$ (or a regular timelike center for
$\alpha=1$). We then consider in Sect.\ 3 the limit $\Lambda\to0$,
which is relevant to the behavior of the corresponding $\Lambda<0$
solutions near the spacelike singularity. The behavior of the
solutions below the threshold ($\alpha>0$) agrees qualitatively with
the results of the numerical simulations of \cite{PC}. Finally we
address in Sect.\ 4 the crucial issue of the linear perturbations of
the $\alpha=0$ solution. Enforcing suitable boundary conditions, we
find only one growing mode, corresponding to the small $\alpha$
solution, as required for the critical solution. In the closing
section we summarize our results and discuss the value of the
critical exponent.

\setcounter{equation}{0}
\section{A class of separable solutions}
The Einstein-scalar field equations are
 \be\lb{mod}
\R_{\mu\nu} - \frac12\R g_{\mu\nu} + \Lambda g_{\mu\nu} =
\kappa\left[\partial_{\mu}\phi\partial_{\nu}\phi - \frac12
g_{\mu\nu}\partial^{\lambda}\phi\partial_{\lambda}\phi\right]\,,
\quad D^{\lambda}\partial_{\lambda}\phi = 0\,,
 \ee
where $\Lambda$ is the cosmological constant, and $\kappa$ the
Einstein gravitational constant, set to one in the following. It was
shown in \cite{crit2} that, in three spacetime dimensions, the
ansatz
 \be\lb{met2}
ds^2 = F^2(T)\left[- dT^2 + dR^2 + G^2(R)d\theta^2\right]\,, \quad
\phi = \phi(T)
 \ee
leads to a solution of the field equations, provided the functions
$G(R)$, $F(T)$ and $\phi(T)$ solve the differential equations
 \ba
&& G^{'2} - \gamma G^2 = \alpha\,, \lb{first2}\\
&& \dot{F}^{2} - \gamma F^2 = \Lambda F^4 + b^2\,, \lb{mast2}\\
&& \dot{\phi} = \sqrt2bF^{-1}\,, \lb{phi2}
 \ea
where $\dot{} = \partial/\partial_T$, $' =\partial/\partial_R$, and
$\gamma$, $\alpha$ and the scalar field strength $b$ are real
integration constants. A not immediately obvious consequence of this
ansatz is the lack of an asymptotic AdS region (see below). The
Ricci scalar of the metric (\ref{met2}) is
 \be\lb{riscal}
{\cal R} = 6\Lambda - \frac{2b^2}{F^4(T)}\,.
 \ee
This may be used to show that this metric has a vanishing Cotton
tensor and so is conformally flat, which also follows from the
observation that it is conformal to a static metric with constant
curvature spatial sections. For $F^2=0$ the conformal factor
vanishes and the metric is singular. Solutions with a regular center
correspond to the initial conditions $G(0)=0$, $G'(0)=1$, and thus
necessarily $\alpha=+1$. In \cite{crit2}, these regular solutions
were only presented as extensions of the Garfinkle solutions to
which they reduce when the cosmological constant is switched off,
but their properties, or those of the more general $\alpha\neq1$
separable solutions, were not further discussed.

Here we briefly discuss the apparent horizon of these separable
solutions, before focussing on the global structure of the
$\Lambda<0$, $\gamma > 0$ solutions. The location $r=r_{AH}$ of the
apparent horizon of the metric (\ref{met2}) (with $r = F(T)G(R)$ the
areal radius) is defined as the solution of the equation
 \be
g^{\mu\nu}\partial_{\mu}r\partial_{\nu}r =
G^2\left[-\frac{\dot{F}^2}{F^2} + \frac{G'^2}{G^2}\right] = 0\,.
 \ee
On account of (\ref{first2}) and (\ref{mast2}), this reduces to
 \be\lb{ah2}
\frac{\alpha}{G^2(R)} = \frac{b^2}{F^2(T)} + \Lambda
F^2(T)\,.
 \ee
The slope of the apparent horizon is given by
 \be\lb{slope}
\left.\frac{dR}{dT}\right|_{AH} = \frac{\dot{F}}{G'}\frac{dG}{dF} =
\pm\frac{F}{G}\frac{dG}{dF} = \mp F^2G^2\frac{dG^{-2}}{dF^2} =
\pm\frac{b^2 - \Lambda F^4}{b^2 + \Lambda F^4}
 \ee
for $\alpha\neq0$. The apparent horizon, if it exists, is for
$\alpha\neq0$ everywhere null for $b^2 = 0$ (vacuum) as well as for
$\Lambda=0$, and is null for $\Lambda\neq0$, $F^2(T) = 0$. It is
spacelike for $\alpha=0$, as well as for $b^2\neq0$, $\alpha\neq0$,
and $\Lambda<0$.

To understand how these solutions might describe gravitational
collapse, it is important to know first how the background vacuum
($b^2=0$) spacetime is described by (\ref{met2}). For $\Lambda<0$,
solutions to (\ref{mast2}) with $b^2=0$ exist only if $\gamma > 0$,
which we assume in the following, choosing without loss of
generality $\gamma=+1$. We first consider the case $\alpha<0$.
Putting $\Lambda=-l^{-2}$ and $\alpha=-M$, the solution of equations
(\ref{first2}) and (\ref{mast2}) is $G=\sqrt{M}\cosh R$, $F=l/\cosh
T$, so that the metric (\ref{met2}) reads
 \be\lb{btzd}
ds^2 = \frac{l^2}{\cosh^2(T)}\left[- dT^2 + dR^2 +
M\cosh^2(R)d\theta^2\right]\,.
 \ee
The spacetime is constant curvature and with a null bifurcate
apparent horizon $R \pm T = 0$, so the metric (\ref{btzd}) must be
(locally) isometric to that of BTZ. Indeed, performing on the static
BTZ metric
 \be\lb{btz}
ds^2 = -(r^2/l^2-M)\, dt^2+\frac{dr^2}{r^2/l^2-M} + r^2\,d\theta^2
 \ee
the coordinate transformation
 \be
\frac{r}{l\sqrt{M}} = \frac{\cosh R}{\cosh T}\,, \quad
\coth\left(\frac{\sqrt{M}\,t}{l}\right) = \left(\frac{\sinh R}{\sinh
T}\right)^{\epsilon}\,,
 \ee
with $\epsilon = {\rm sgn}(r^2-Ml^2)$, leads to the metric
(\ref{btzd}). Note however that the metric (\ref{btzd}) covers only
the domain
 \be\lb{domain}
\tanh^2\left(\frac{\sqrt{M}\,t}{l}\right) < \frac{r^2}{Ml^2} <
\coth^2\left(\frac{\sqrt{M}\,t}{l}\right)
 \ee
of the maximally extended static BTZ spacetime. This domain, bounded
by the coordinate singularity $T\to\infty$ of the metric
(\ref{btzd}), does not contain spacelike infinity $r\to\infty$, i.e.
the metric (\ref{btzd}) is not asymptoitically AdS.

Similarly, the separable metric for $\alpha=0$ ($G= e^R$)
 \be
ds^2 = \frac{l^2}{\cosh^2(T)}\left[- dT^2 + dR^2 +
\e^{2R}d\theta^2\right]
 \ee
is locally transformed into the Kruskal form of the BTZ vacuum
($M=0$) metric \cite{BTZ}
 \be\lb{vacd}
ds^2 = \frac{l^2}{\rho^2}\left[- d\tau^2 + d\rho^2 +
d\theta^2\right]
 \ee
($\rho=-l/r$) by the coordinate transformation
 \be\lb{exptrans}
\tau = \e^{-R}\,\sinh T\,, \quad \rho = -\e^{-R}\,\cosh T\,.
 \ee
Finally, the metric (\ref{met2}) with $\alpha>0$
($G=\sqrt{\alpha}\sinh R$) can be transformed into that of (a domain
of) $AdS_3$ with a conical singularity (unless $\alpha=+1$) by the
coordinate transformation
 \be
\frac{r}{l\sqrt{\alpha}} = \frac{\sinh R}{\cosh T}\,, \quad
\cot\left(\frac{\sqrt{\alpha}\,t}{l}\right) = \frac{\cosh R}{\sinh
T}\,.
 \ee

Now we turn to the non-trivial solutions with a dynamical scalar
field, corresponding to $b^2\neq0$. The metric (\ref{met2}) for
these solutions may be written \cite{crit2} in the explicit
form\footnote{Another, related, FRW explicit form is given in
Appendix A.}
 \be\lb{etamet}
ds^2 = - \frac{l^2}4\,\frac{d\eta^2}{(\eta-\eta_-)(\eta_+-\eta)} +
\eta\,[ dR^2 + G^2(R)\,d\theta^2]\,,
 \ee
with $\eta=F^2(T)$, and
 \be
\eta_{\pm} = \frac{l}2\left[l \pm \sqrt{l^2 + 4b^2}\right]\,,
 \ee
with $\eta_-<0\le\eta\le\eta_+$. The spacetime is time-symmetric
around the turning point $\eta=\eta_+$, with two symmetrical past
and future central spacelike singularities $\eta=0$.

For $\alpha=-M<0$ the spacetime is also, as in the vacuum case,
left-right symmetric. Writing the equation of the apparent horizon
(\ref{ah2}) as $\eta=\eta_{AH}(R)$, we see that $bl < \eta_{AH}(R)
\le \eta_+$, the time-symmetric value $\eta_{AH}=\eta_+$
corresponding to the space-symmetric value $G=\sqrt{M}$. It follows
that, as in the case of the BTZ black hole, the apparent horizon is
bifurcate and divides the spacetime into two left and right exterior
regions $I$ and $I'$, and two future and past interior regions $II$
and $II'$.  However, to the difference of the BTZ black hole, the
apparent horizon is spacelike. Another remarkable difference is
that, similarly to the case of its vacuum counterpart (\ref{btzd}),
the metric (\ref{etamet}) is not asymptotically AdS. The
corresponding spacetime is bounded by the spacelike singularity
$F^2=0$, which for $b^2\neq0$ is a genuine curvature singularity.
Near this singularity the metric (\ref{met2}) can be approximated,
from (\ref{mast2}), by
 \be\lb{apprmet}
ds^2  \simeq  F^2[-b^{-2}\,dF^2 + dR^2 + G^2(R)\,d\theta^2]\,,
 \ee
so that the two-dimensional sections are conformal to Minkowski$_2$,
the two components of the singularity $\eta \equiv F^2 = 0$
constituting the boundary of the spacetime (Fig. 1a).

Because asymptotic infinity is only a point, this solution does not
have an event horizon. It does have in common with the static BTZ
black hole a bifurcate apparent horizon. Because of this topological
similarity, we will refer to this dynamical solution as ``black-hole
like''. By analogy with the case of the BTZ black hole, we introduce
the mass function
 \be\lb{mass}
M(\eta_{AH}) \equiv \frac{r^2_{AH}}{l^2}\,.
 \ee
In the case of the BTZ black hole ($b^2=0$), the apparent horizon
coincides with the event horizon $r^2=Ml^2$, so that
$M(\eta_{AH})=M$. For $b^2\neq0$,
 \be
M(\eta_{AH}) = \frac{M\eta_{AH}^2}{\eta_{AH}^2-b^2l^2} \ge
\frac{M\eta_+}{l^2} > M\,.
 \ee
We define the effective black hole mass as the minimum value of
$M(\eta_{AH})$,
 \be\lb{Meff}
M_{eff} = \frac{M\eta_+}{l^2}\,,
 \ee
which is attained at the moment of time symmetry. For small scalar
field strengths, $b^2 \ll l^2$, $M_{eff} \simeq M(1+b^2/l^2)$, with
$Mb^2/l^2$ the scalar field contribution to the black hole mass,
while for large scalar field strengths, $M_{eff} \simeq Mb/l$.

For $\alpha=0$, $\eta_{AH} = bl < \eta_+$, so that the solution,
given by (\ref{etamet}) with $G=e^R$ has an apparent horizon with
two disjoint components, dividing the spacetime into two future and
past interior regions $II$ and $II'$ and a single exterior region
$I$ (Fig. 1b). The minimum value of the mass function
$M(\eta_{AH})=bl^{-1}\e^{2R}$ is now zero. This solution corresponds
to that discussed in \cite{BST}. In this case, the coordinate
transformation (\ref{exptrans}) transforms the separable metric
(\ref{met2}) into the explicitly conformally flat metric \cite{BST}
 \be\lb{bst}
ds^2 = \frac{F^2(T)}{\rho^2-\tau^2}\left[- d\tau^2 + d\rho^2 +
d\theta^2\right] \quad (\tanh T = - \tau/\rho)
 \ee
(an explicit form of the function $F^2(T)$ in terms of elliptic
functions is given in \cite{BST}). In the limit $b^2\to0$ this
reduces to the BTZ vacuum metric (\ref{vacd}).

For $\alpha>0$, $\eta_{AH} < bl < \eta_+$, and again there is an
apparent horizon with two disjoint components. However the spacetime
boundary has now a third component, the center $G=0$ which is
regular if $\alpha =+1$ (Fig. 1c).

\setcounter{equation}{0}
\section{The sub-threshold solution near the singularity}

We wish to explore whether the $b^2\neq0$, $\alpha=0$ solution,
which lies at the threshold for $\alpha<0$ scalar field black-hole
like solutions and reduces to the BTZ vacuum for $b^2\to0$, presents
the properties required of the critical solution for scalar field
collapse in $AdS_3$. To this end, we shall compare the behavior of
the sub-threshold $\alpha>0$ solution near the spacelike curvature
singularity with that observed in the numerical simulations of
\cite{PC}. Near the singularity, the cosmological constant can be
neglected, so that the function $F(T)$ can be approximated by the
solution of (\ref{mast2}) with $\Lambda=0$, $F = b\sinh{T}$, while
the general solution of (\ref{first2}) can be written, up to a
translation, as $G = \e^R - (\alpha/4)\e^{-R}$, and the family of
near-threshold solutions near the singularity $T=0$ can be written
 \ba\lb{L0}
ds_0^2 &=& b^2\sinh^2(T)\left[- dT^2 + dR^2 + \left(\e^R -
\frac{\alpha}4\e^{-R}\right)^2\,d\theta^2\right]\,, \nn\\
\phi_0 &=& \sqrt2\,\ln\tanh(-T/2)
 \ea
(we assume $T<0$). The metric (\ref{L0}) can be put in the double
null form
 \be\lb{dnull}
ds^2 = \e^{2\sigma}\,du\,dv + r^2\,d\theta^2
 \ee
by the coordinate transformation
 \be\lb{cuv0}
u = b\,\e^{R-T}\,, \quad v = b\,\e^{R+T}
 \ee
($u>v>0$), leading to
 \be\lb{uva+}
ds_0^2 = \frac{(u-v)^2}4\left[b^2\frac{dudv}{(uv)^2} +
\left(1-\frac{\beta^2}{4uv}\right)^2d\theta^2\right]\,, \quad \phi_0
= \sqrt2\,\ln\left(\frac{u^{1/2}-v^{1/2}}{u^{1/2}+v^{1/2}}\right)\,,
 \ee
with $\beta^2 = \alpha b^2$.

A necessary condition for the $\alpha=0$ solution
$(\ol{\sigma}_0,\ol{r}_0,\ol{\phi}_0)$ to qualify as a possible
critical solution is that it be self-similar. Let us first show that
this is indeed the case. For $\Lambda=0$ the solution of the
Einstein equation $r_{uv}=0$ can be gauge transformed to $r_0 =
(u-v)/2$. Inserting the self-similar ansatz \cite{G}
 \be
\phi = c\ln{u} + \psi(y)\,,
 \ee
with $y^2=v/u$, into the scalar field equation
 \be
2r\phi_{,uv} + r_{,u}\phi_{,v} + r_{,v}\phi_{,u} = 0
 \ee
we obtain the general solution (up to an additive constant)
 \be\lb{phiback}
\phi = c\ln u + (c+d)\ln(1-y) + (c-d)\ln(1+y)\,,
 \ee
with $c$ and $d$ two independent constants\footnote{This solution
was previously given in \cite{crit2}, Eq. (2.14).}. The assumption
of a regular center at $r=0$ ($y=\pm1$) leads to Garfinkle's
solutions with $c=\mp d$ \cite{G}, while the solution (\ref{uva+})
for $\alpha=0$ corresponds to $c=0$, $d^2=2$.

For $\alpha>0$ the spacetime structure is more obvious in the
alternate double-null coordinates $\ol{u}$, $\ol{v}$ defined by
 \ba\lb{cuva+}
\ol{u} &=& u+\frac{\beta^2}4u^{-1} = \beta\cosh(\ol{R}-T) \,,\nn\\
\ol{v} &=& v+\frac{\beta^2}4v^{-1} = \beta\cosh(\ol{R}+T)\,,
 \ea
with $\ol{R} = R - (1/2)\ln(\alpha/4)$. These coordinates are such
that $r =(\ol{u}-\ol{v})/2$, which vanishes either for $\ol{R} = 0$
(timelike center, corresponding to a conical singularity if
$\alpha\neq1$) or $T = 0$ (spacelike curvature singularity). In the
sector ($\ol{R} > 0$, $T<0$) the apparent horizon is at $T =
-\ol{R}$, corresponding to $\ol{v} = \beta$ ($v = \beta/2$),
dividing the spacetime in two sectors $I$ and $II$. Noting that
$d\ol{v}/dv= 1-\e^{-2(R+T)}$ is negative in sector $I$ (the sector
of the center) and positive in sector $II$ (the sector of the
curvature singularity), we obtain the form of the metric
 \be\lb{uva+0}
ds_0^2 = \mp\frac{\ol{u}\,\ol{v}\pm
\sqrt{(\ol{u}^2-\beta^2)(\ol{v}^2-\beta^2)} -
\beta^2}{2\alpha\sqrt{(\ol{u}^2-\beta^2)(\ol{v}^2-\beta^2)}}d\ol{u}d\ol{v}
+ \frac{(\ol{u}-\ol{v})^2}4\,d\theta^2\,,
 \ee
with the upper sign in sector $I$ and the lower sign in sector $II$.

For comparison with the numerical simulations, we define a time
coordinate $\hat{u}$ (noted $t_c$ in \cite{PC} and $u$ in \cite{G})
as the null coordinate which coincides with the proper time of the
central observer, measured from the curvature singularity. At the
center, $\ol{v}=\ol{u}$, so that (\ref{uva+0}) reduces to
 \be
-d\hat{u}^2 = -\alpha^{-1}\,d\ol{u}^2\,,
 \ee
leading to
 \be
\hat{u} = \alpha^{-1/2}(\beta-\ol{u}) = b[1-\cosh(\ol{R}- T)]\,.
 \ee
Putting $\hat{u} = -\e^{-\hat{T}}$ and defining a relative radial
coordinate $\hat{R}$ by
 \be\lb{hatR}
\hat{R} = r\,\e^{\hat{T}} =
\frac{\alpha^{1/2}}2\,\frac{\ol{u}-\ol{v}}{\ol{u}-\beta}\,,
 \ee
we show in Appendix B that, assuming $\hat{T}$ large, the scalar
field is given in sector I by
 \be\lb{phiG}
\phi = \sqrt2\left(-\frac12\,\hat{T} + \ln\left[\frac{1 +
\sqrt{1-2\alpha^{-1/2}\hat{R}}}2\right] + O(\e^{-\hat{T}})\right)\,.
 \ee
We note that, to this approximation, this coincides with Eq. (18) of
\cite{G} (where $\kappa$ has been set to $4\pi$) for the $n=1$
Garfinkle solution, provided the radial coordinate $R$ in Eq. (18)
is replaced by $\alpha^{-1/2}\hat{R}$.

Following \cite{PC}, we define the logarithmic coordinates
 \be
Z = \ln r = \ln\hat{R} - \hat{T}\,, \quad \hat{Z} = Z -
\frac12\ln\alpha\,,
 \ee
and compute
 \be\lb{phizz}
\phi_{,ZZ}(Z,\hat{T}) = -
\frac1{\sqrt2}\,\frac{\e^{\hat{Z}+\hat{T}}}{(1-2\e^{\hat{Z}+\hat{T}})^{3/2}}
+ O(\e^{-\hat{T}})\,.
 \ee
At a fixed $\hat{T}$, this quantity depends on $Z$ only through the
combination $Z - (1/2)\ln\alpha$, in accordance with Fig. 17 of
\cite{PC}, which shows to a good approximation that when a point
particle of mass $1-\e^{-2A(0,0)}$ (a deficit angle $\omega =
2\pi(1-\e^{-A(0,0)}$)) is introduced at the center, $\phi_{,ZZ}$
depends only on $Z$ through the difference $Z - A(0,0)$. The fact
that $\phi_{,ZZ}$ depends only on the sum $Z+\hat{T}$ is also in
good agreement with Fig. 10 of \cite{PC}.

We can also evaluate the Ricci scalar (\ref{riscal}). From
(\ref{riscalA}) this diverges at the origin as
 \be
{\cal R}(0) \simeq -\frac12\e^{2\hat{T}}\,,
 \ee
in accordance with \cite{PC}, where it was found that the value of
the curvature scalar $R$ at the origin diverges like $1/t_c^2$ as
one approaches the accumulation point. The combination
 \be
r^2\,{\cal R} \simeq - 2\alpha\left(\frac{1-\e^{\hat{Z}+\hat{T}}
-\sqrt{1-2\e^{\hat{Z}+\hat{T}}}}{\e^{\hat{Z}+\hat{T}}}\right)^2 \,.
 \ee
depends only on $Z+\hat{T}$ and goes to zero for $Z\to-\infty$ ($T$
fixed) as $-(\alpha/2)\e^{2(\hat{Z}+\hat{T})}$, in agreement with
Fig. 12 of \cite{PC}. The mass aspect
 \be
M(R,T) = \frac{r^2}{l^2} - g^{\mu\nu}\partial_{\mu}r\partial_{\nu}r
= \frac{b^2G^2}{F^2} - \alpha \simeq - \frac{r^2\,{\cal R}}2 -
\alpha
 \ee
similarly depends only on $\hat{Z}+\hat{T}$ and goes to $-\alpha$
for $Z\to-\infty$ at $T$ fixed, in agreement (in the case
$\alpha=1$) with Fig. 11 of \cite{PC}.

We have already noted that, without the higher-order corrections,
the scalar field of Eq. (\ref{phiG}) is exactly that
 which would have been obtained by the self-similar ansatz
of \cite{G} for the parameter $n=1$ if an angular deficit $\omega =
2\pi(1-\alpha^{1/2})$ had been introduced in the metric ansatz.
However for $\alpha=1$ this does not agree with Fig. 2 of \cite{G},
which is well fitted by $n=4$. To compare with Fig. 1 of \cite{G},
we can evaluate exactly
 \be\lb{fig1g}
\left.\frac{\partial\phi_0}{\partial\hat{T}}\right\vert_{r=0} =
\left.\frac{\sqrt2}{\sinh T}\,\frac{\partial T}{\partial\hat{T}}
\right\vert_{r=0} = - \frac1{\sqrt2}\left[1+\frac{e^{-\hat
T}}{2b}\right]^{-1} \,,
 \ee
which goes to the limit $-1/\sqrt2$ when $\hat{T} \to \infty$. In
comparison, Fig. 1 of \cite{G} for
$\partial\phi_0/\partial\hat{T}\vert_{r=0}$ shows a constant plateau
at approximately $-0.264$, corresponding to a value of the Garfinkle
parameter $n=4$, followed by an increase which might be consistent
with the limiting value $-1/\sqrt{8\pi} \simeq -0.199$. However
(after rescaling our scalar field by $1/\kappa$ with
$\kappa=\sqrt{4\pi}$), our (\ref{fig1g}) decreases rather than
increases towards this value.

\setcounter{equation}{0}
\section{Perturbations}

Another property required of a critical solution is that it has only
one unstable growing mode under linear perturbations. To study the
linear perturbations of the $\alpha=0$ solution, (\ref{etamet}) with
$G=e^R$, we follow the procedure described in \cite{CCF}. Taking as
independent variables $u$ and $y=(v/u)^{1/2}$, the Einstein and
scalar field equations (\ref{mod}) may be rewritten as
 \ba
&&2r_{uy} - u^{-1}(yr_{yy} + r_y) = 2\Lambda uyr\e^{2\sigma}\,, \lb{Ey1} \\
&&4\sigma_{uy} - 2u^{-1}(y\sigma_{yy} + \sigma_y) = 2\Lambda
uy\e^{2\sigma} + \phi_y(-2\phi_u
+ u^{-1}y\phi_y)\,, \lb{Ey2} \\
&&yr_{yy} - r_y - 2y\sigma_yr_y = - ry\phi_y^2\,, \lb{Ey3} \\
&&- r_{uu} + u^{-1}yr_{uy} - u^{-2}yr_y + 2\sigma_ur_u - u^{-1}y(
\sigma_ur_y + \sigma_yr_u) = \nn\\
&&\qquad \qquad \qquad = r(\phi_u^2 - u^{-1}y\phi_u\phi_y)\,, \lb{Ey4} \\
&&r[2\phi_{uy} - u^{-1}(y\phi_{yy} + \phi_y)] + r_u\phi_y +
r_y\phi_u - u^{-1}yr_y\phi_y  = 0\,. \lb{Sy}
 \ea
The $\alpha=0$ solution is put in the double-null form (\ref{dnull})
by the coordinate transformation (\ref{cuv0}). A generic solution
$(r,\phi,\sigma)$ of the field equations is linearized around this
background solution $(\ol{r},\ol{\phi},\ol{\sigma})$ as
 \ba\lb{lin}
r &=& \ol{r}(u,y) + \varepsilon u^{1-k}f(y)\,, \nn\\
\phi &=&  \ol{\phi} + \varepsilon u^{-k}\sqrt2\,g(y)\,, \nn\\
\sigma &=& \ol{\sigma} + \varepsilon u^{-k}h(y) \,,
 \ea
with $k$ a real constant. Modes with Re$(k) > 0$ grow as the
singularity $u = 0$ is approached. To separate the linearized field
equations, both the background solution and its linear perturbations
are expanded in powers of the dimensionless quantity $\Lambda u^2$,
so that to each order the linearized field equations reduce to
ordinary differential equations in the variable $y$. These
differential equations are solved iteratively, order by order. Here
we shall only treat the problem to zeroth order, with the
$\Lambda=0$ solution (\ref{uva+}) with $\beta=0$ as background,
 \ba\lb{back}
\ol{r}_0 &=& u\frac{1-y^2}2 \,, \nn\\
\ol{\phi}_0 &=& \sqrt2\,\ln\frac{1-y}{1+y} \,, \nn\\
\ol{\sigma}_0 &=& \ln(b/2) - \ln u + \ln(1-y^2) - 2\,\ln y \,.
 \ea

Taking $F = b\sinh{T}$ in (\ref{bst}) leads to the alternate form of
the truncated background metric
 \be\lb{rt0}
ds_0^2 = \frac{b^2\tau^2}{(\rho^2-\tau^2)^2}\left(-d\tau^2 + d\rho^2
+ d\theta^2\right) \quad (0<-\tau<\rho)\,.
 \ee
The structure of this spacetime may be visualized in the plane
$(x,\tau)$, where $x \equiv 1/\rho$. Noting that
 \be
x\tau = \frac{y^2-1}{y^2+1}\,,
 \ee
we see that the curves $y=$ constant are hyperbolas bounded on one
side by $x\tau=-1$ ($y=0$), on the other side by $x\tau=0$ ($y=1$).
The Penrose diagram of the truncated background is thus a triangle
bounded by the regular timelike line $\rho=\infty$ ($R=\infty$), the
singular spacelike line $\tau=0$ ($T=0$), both components of the
boundary $y=1$, and the null line $y=0$ ($T=-\infty$). The
intersections of the null line $y=0$ with the timelike line
$\rho=\infty$ and the spacelike line $\tau=0$ correspond
respectively to $u=\infty$ and $u=0$. Actually, the null boundary
$T=-\infty$, which is at infinite geodesic distance, is the limit
$l\to\infty$ of the apparent horizon $F_{AH}^2=bl$ of the spacetime
(\ref{bst}), so that our truncated background is entirely contained
in the future region $II$ of the full background (Fig. 1b).

The linearization of (\ref{Ey1}), (\ref{Sy}) and (\ref{Ey3}) with
$\Lambda=0$ around the truncated background (\ref{back}) yields the
differential equations
 \ba
&& yf_0'' + (2k-1)f_0' = 0\,, \lb{eqf}\\
&& y(1-y^2)g_0'' + 2[k - (k+1)y^2]g_0' - 2kyg_0 = \nn \\
&& \qquad \qquad \frac{4y}{1-y^2}f_0' +
\frac{4[k+(2-k)y^2]}{(1-y^2)^2}f_0\,, \lb{eqg}\\
&& 2y(yh_0'-2g_0') = -yf_0'' - \frac{3+y^2}{1-y^2}f_0' -
\frac{8y}{(1-y^2)^2}f_0\,, \lb{eqh}
 \ea
and the linearization of (\ref{Ey2}) and (\ref{Ey4}) yields the
extra (constraint) equations
 \ba
&& yh_0'' + (2k+1)h_0' - \frac{4(yg_0'+kg_0)}{1-y^2} = 0\,,\lb{cons1}\\
&& \frac{y(1-y^2)}2\,h_0' + kh_0 - 2kyg_0 = \nn\\
&& \qquad \qquad
-(k-1)\left[yf_0'+\left(k-2+\frac2{1-y^2}\right)f_0\right]\lb{cons2}\,.
 \ea
To solve these equations one must enforce appropriate boundary
conditions. The boundary of our truncated background (\ref{rt0}) has
two regular components.  Our first boundary condition shall be that
the perturbation be regular on the timelike boundary $\rho\to\infty$
($y=1$). On the other regular boundary component, $y=0$, we shall
enforce the condition that the perturbation be, as the background
(\ref{back}), analytic in $y$. The rationale for this condition is
that, as we have seen, this null boundary of the truncated
background is actually an apparent horizon of the full background,
so that this analyticity condition ensures that the perturbation can
be extended from region $II$ to region $I$ of the full background.

To discuss the solution of the system (\ref{eqf})-(\ref{eqh}) we
follow the procedure of \cite{GG}. The general solution of
(\ref{eqf}) is
 \be\lb{fgen}
f_0 = c_0 + c_1(1-y^{2-2k})
 \ee
(for $k=1$, a linearly independent solution is $c'_1\log{y}$, which
can be excluded by the analyticity condition). This is actually a
pure gauge perturbation, generated from the background $\ol{r}_0 =
u(1-y^2)/2$ by the gauge transformation
 \be
u \rightarrow u + 2\varepsilon(c_0+c_1)u^{1-k}\,, \quad v\rightarrow
v + 2\varepsilon c_1 v^{1-k}\,,
 \ee
corresponding to
 \be\lb{gauge}
\delta y = \varepsilon u^{-k}y[c_1y^{-2k} - (c_0+c_1)]\,.
 \ee
In the gauge $c_0=c_1=0$, equation (\ref{eqg}) is homogeneous. Its
solution which is regular in $y = 1$ is
 \be
g_0 = c_2F(k,1/2,1;1-y^2)\,,
 \ee
with $F$ a hypergeometric function. To this we must add a particular
solution of the inhomogeneous equation given by the gauge
transformation (\ref{gauge}) acting on the background solution
 \be
g_{(g)} = [c_0 + c_1(1-y^{-2k})]\frac{2y}{1-y^2}\,.
 \ee
This is certainly not regular in $y=1$ if $c_0\neq0$. Taking from
now on $c_0=0$, and using an identity between hypergeometric
functions, we obtain the solution of (\ref{eqg}) in the gauge $c_1$
 \ba\lb{ggen}
g_0 &=& c_1\frac{2y(1-y^{-2k})}{1-y^2} +
\frac{c_2\Gamma(1/2-k)}{\sqrt\pi\,\Gamma(1-k)}\,F(k,1/2,k+1/2;y^2) +
\nn\\
&& \qquad +
\frac{c_2\Gamma(k-1/2)}{\sqrt\pi\,\Gamma(k)}\,y^{1-2k}\,F(1-k,1/2,3/2-k;y^2)
\,,
 \ea
except if $k=(2p+1)/2$ with $p$ integer. In this last case the
transformation between hypergeometric functions of $1-y^2$ and of
$y^2$ involves a non-analytic $\ln y$ term, so that the condition of
analyticity at $y=0$ dictates $c_2=0$, and the solution is pure
gauge. Other non-integer values of $k$ may similarly be eliminated
because the third term is the product of an infinite even series by
the non-analytic monomial $y^{1-2k}$.

There only remain the positive integer values $k=n$. For these
values, the second term in (\ref{ggen}) is identically zero, while
the hypergeometric function in the third term is a polynomial of
order $2k-2$. Expanding in powers of y, we obtain
 \be
g_0 = -2c_1y^{1-2k}\left[1 + y^2 + \cdots + y^{2k-2}\right] +
c_2y^{1-2k}\left[1 + \frac{1-k}{3-2k}y^2 + \cdots +
O(y^{2k-2})\right]\,.
 \ee
The leading divergence is compensated in the gauge $c_2=2c_1$. For
$k=1$, the compensation is exact, $g_0=0$ in this gauge, $f_0=0$ by
virtue of (\ref{fgen}), and $h_0=0$ from (\ref{eqh}) and
(\ref{cons2}), so that the perturbation vanishes altogether. For
$k>1$ there remains
 \be
g_0 = -2c_1y^{3-2k}\left[\frac{k-2}{2k-3} + \cdots +
O(y^{2k-4})\right]\,.
 \ee
For $k=2$ the sum contains only one term, which vanishes
identically, so that the analytic solution in the gauge $c_2=2c_1$
is $g_0=0$. For $k>2$, the solution is not analytic.

It remains to determine the function $h_0(y)$. For $k=2$,
$f_0=c_1(1-y^{-2})$ which annihilates the right-hand sides of
(\ref{eqg}), (\ref{eqh}) and (\ref{cons2}). Eq. (\ref{eqh}) with
$g_0=0$ then leads to $h_0'=0$, and finally $h_0=0$ from
(\ref{cons2}). The solution for $k=2$ is thus
 \be
f_0=c_1(1-y^{-2})\,, \quad g_0 = 0\,, \quad h_0 = 0\,.
 \ee
The resulting linearized solution (\ref{lin}) is identical, with the
correspondence $\varepsilon c_1 = \alpha b^2/8$, to the exact
near-threshold solution (\ref{uva+}) with $\alpha\neq0$.

\section{Discussion}
We have proposed an alternative scenario for circularly symmetric
scalar field collapse in $AdS_3$, in which the critical solution is
self-similar but, instead of having a regular timelike center, is
centerless. This scenario is motivated by the observed independence
of the near-critical regime on the angle deficit of the spacetime.
To illustrate this scenario, we have presented a family of exact,
dynamical solutions to the three-dimensional Einstein-scalar field
equations with negative cosmological constant $\Lambda=-l^{-2}$.
These solutions depend on two parameters, a mass parameter $M =
-\alpha$, and a parameter $b$ measuring the scalar field strength.
For $b=0$ these solutions reduce locally to the static BTZ family of
vacuum solutions. For $b\neq0$ and $\alpha<0$, the dynamical
solution is black-hole like, with a global structure topologically
similar to that of the BTZ black hole, and a finite effective mass.
This effective mass vanishes for the $\alpha=0$ solution, which is
self-similar and centerless. The solutions below the black hole
threshold, $\alpha>0$, have a deficit angle $2\pi(1-\alpha)$. We
have discussed the near-singularity behavior of these solutions and
shown that it agrees qualitatively with that observed in numerical
simulations of subcritical collapse. Finally, we have analyzed in
the $\Lambda=0$ approximation the linear perturbations of the
centerless threshold solution, $\alpha=0$. Assuming reasonable
boundary conditions, we found that it has only one unstable growing
mode, as expected for the critical solution.

The realization of the no-center scenario discussed here presents
some shortcomings. A first problem is that our candidate critical
and subcritical solutions are not asymptotically AdS, while one
would expect the true critical solution for scalar field collapse in
$AdS_3$ to be asymptotically AdS. A second, perhaps related, problem
concerns the quantitative comparison of these solutions, which
behave like the $n=1$ Garfinkle solution near the singularity, with
the numerical simulations, which are best fit by the $n=4$ Garfinkle
solution \cite{G}. Another problem is that subcritical solutions
should be regular, and the critical solution should not have trapped
surfaces, while our threshold and sub-threshold solutions both
present apparent horizons and spacelike singularities. However this
is also the case for the Garfinkle critical solutions (the
corresponding subcritical solutions are not known in closed form,
and therefore one cannot exclude that the same features would be
present there as well), which also present an apparent horizon and,
for $n$ odd, a spacelike singularity \cite{crit}. As in the
Garfinkle case \cite{G}, one should not expect the numerical
critical solution to approach our exact self-similar critical
solution outside the past light cone of the singularity $u=0$. A
fortiori we would not expect our exact $\alpha=1$ subthreshold
solutions, which are regular at the origin but not asymptotically
AdS, to reproduce all the features of the observed near-critical
behavior.

A last problem concerns the value of the critical exponent $\gamma$.
This is defined by the scaling relation of the effective black hole
mass $M_{eff} \propto |p-p^{*}|^{2\gamma}$, and is related to the
mode eigenvalue $k$ by $\gamma=1/k$. Our family of near-critical
solutions depends on the parameter $\alpha=-M$ with the critical
value $\alpha^{*}=0$, so that (\ref{Meff}) leads to the value
$\gamma=1/2$. This agrees with the value obtained in our linear
perturbation analysis, $k=2$, corresponding to a critical exponent
$\gamma=1/2$. However this value disagrees with the numerical
evaluations of \cite{PC} from maximum central curvature scaling,
leading to $\gamma=1.20\pm 0.05$, and of \cite{HO} from apparent
horizon mass scaling, which yield $\gamma\simeq0.81$.

An interesting collateral result of our investigation concerns the
existence of black holes in three-dimensional spacetime. It is often
taken for granted that the only black hole solutions to the
three-dimensional Einstein-scalar field system are the BTZ black
holes, with negative cosmological constant and vanishing scalar
field. Actually, two non-trivial counterexamples with zero
cosmological constant are already known: the cold black holes of
\cite{sig} with a negative gravitational constant; and (in the case
of a positive gravitational constant) the Garfinkle solutions with
$n$ odd. In the present work, we have presented a new dynamical
scalar-field black-hole like solution, with a spacelike singularity
hidden behind a spacelike apparent horizon, corresponding to the
over-threshold case $\Lambda<0$ and $\alpha \le 0$, with
$M_{eff}\ge0$.

In conclusion, we have presented a scenario for critical scalar
field collapse in $AdS_3$ alternative to that of \cite{G} which,
despite a number of shortcomings, has the advantage of explaining
black hole formation, and the independence of the critical behavior
on the angle deficit of the spacetime. Perhaps there is a (yet
unknown) family of asymptotically AdS solutions which reduces to the
BTZ family when the scalar field is switched off, with a
self-similar critical solution devoid of a timelike center, and with
subcritical solutions which would behave near the singularity like
the n = 4 Garfinkle solutions.

\subsection*{Acknowledgment}
We wish to thank D. Garfinkle, C. Gundlach, V. Husain and  F.
Pretorius for informative discussions.

\renewcommand{\theequation}{A.\arabic{equation}}
\setcounter{equation}{0}
\section*{Appendix A: FRW form of the separable solution}
The solution (\ref{met2})-(\ref{phi2}) may also be written in the
Friedmann-Robertson-Walker form
 \ba
ds^2 &=& - d\ol{t}^2 + \frac{l^2}2\left[1 +
\nu\cos\left(\frac{2\ol{t}}l\right)\right]\left[\frac{d\ol{r}^2}{\alpha+\ol{r}^2}
+ \ol{r}^2\,d\theta^2\right]\,,\lb{frw}\\
\phi &=& \frac1{\sqrt2}\,\ln\left[\frac{(1+\nu)\,l/2b +
\tan(\ol{t}/l)}{(1+\nu)\,l/2b - \tan(\ol{t}/l)}\right]
\quad\left(\nu = \sqrt{1+4b^2/l^2}\right)\,, \lb{scalfrw}
 \ea
where $d\ol{t} = F(T)dT$ and $\ol{r} = G(R)$. In the weak-amplitude
limit (no back-reaction) $b^2 \ll l^2$, the metric (\ref{frw})
reduces to the BTZ cosmology
 \be\lb{btzcosm}
ds^2 = - d\ol{t}^2 +
l^2\,\cos^2\left(\frac{\ol{t}}l\right)\left[\frac{d\ol{r}^2}{\alpha+\ol{r}^2}
+ \ol{r}^2\,d\theta^2\right]\,,
 \ee
and the scalar field (\ref{scalfrw}) reduces to the solution of the
wave equation
 \be\lb{scalweak}
\phi = \frac{\sqrt2 b}{l}\,\tan\left(\frac{\ol{t}}{l}\right) =
\frac{\sqrt2 b}{l}\,\sinh{T} = \frac{\sqrt2
b}{l}\,\left(\frac{r^2+\alpha l^2}{\alpha
l^2\cot^2(\sqrt{\alpha}t/l)-r^2}\right)^{1/2}
 \ee
over the BTZ background (\ref{btzcosm}), (\ref{btzd}) or (\ref{btz})
(with $M=-\alpha$), respectively. The scalar field (\ref{scalweak})
diverges at the coordinate singularity $T\to\infty$ of (\ref{btzd})
(or the singularity $\ol{t}=\pi l/2$ of (\ref{btzcosm})).

Noting that in the coordinates of (\ref{frw}) $d\hat{u} = - d\ol{t}$
at $r$ fixed, we can use (\ref{scalfrw}) to evaluate directly
$\partial\phi/\partial\hat{T}|_{r=0}$ without the approximation
$\Lambda=0$, leading to
 \be
\left.\frac{\partial\phi}{\partial\hat{T}}\right\vert_{r=0} = -
\frac1{\sqrt2}\,\frac{\hat{u}}{\sin{\hat{u}}[\cos{\hat{u}} -
(l/2b)\sin{\hat{u}}]} \,,
 \ee
with $\hat{u} = -\e^{-\hat{T}}$. This reduces to (\ref{fig1g}) in
the limit $l\to\infty$, but has a rather different behavior when
$b/l =O(1)$, and for large $b/l$ increases towards the limit
$-1/\sqrt2$ when $\hat{T} \to \infty$.

\renewcommand{\theequation}{B.\arabic{equation}}
\setcounter{equation}{0}
\section*{Appendix B: Near-singularity behavior of the scalar field (\ref{L0})}
We first evaluate the relative radial coordinate $\hat{R}$ of
(\ref{hatR}) in terms of $\hat{T}$ and $\ol{R}$. The numerator is
 \ba\lb{uvmunu}
\ol{u} - \ol{v} &=& \beta[\cosh(\ol{R}-T) - \cosh(\ol{R}+T)] \nn\\
&=& 2\beta\sinh\ol{R}[\sinh(\ol{R}-T)\cosh\ol{R} - \cosh(\ol{R}-T)\sinh\ol{R}] \nn\\
&=& \frac{8\beta\nu(\mu-\nu)(1-\mu\nu)}{(1-\mu^2)(1-\nu^2)^2}\,,
 \ea
where we have put $\mu = \tanh[(\ol{R}-T)/2]$, $\nu =
\tanh[\ol{R}/2]$. We note that
 \be
\e^{-\hat{T}} = b[\cosh(\ol{R}-T) - 1] = \frac{2b\mu^2}{1-\mu^2}\,,
 \ee
so that $\hat{T}$ large means $\mu^2 \ll 1$. Also, in the region I,
$\ol{R}<-T$, implying $\ol{R}<(\ol{R}-T)/2$, hence $\nu<\mu/2$, so
that $\mu\nu$ and $\nu^2$ are also small. So we can approximate
(\ref{uvmunu}) by
 \be
\ol{u} - \ol{v} \simeq 8\beta\nu(\mu-\nu)\,.
 \ee
Similarly, $\ol{u} - \beta = \alpha^{1/2}\e^{-\hat{T}} \simeq
2\beta\mu^2$, so that
 \be
\hat{R} \simeq \frac{2\alpha^{1/2}\nu(\mu-\nu)}{\mu^2} =
2\alpha^{1/2}\,x(1-x)\,,
 \ee
with $x = \nu/\mu < 1/2$. This can be inverted to yield
 \be\lb{xR}
x = \frac{1 - \sqrt{1-2\alpha^{-1/2}\hat{R}}}2\,.
 \ee
Next we compute
 \ba
\phi &=& \sqrt2\,\ln\tanh(-T/2) =
\sqrt2\,\ln\tanh\left[(\ol{R}-T)/2 -T/2\right] \nn\\
&=& \sqrt2\,\ln\left(\frac{\mu-\nu}{1-\mu\nu}\right) \simeq
\sqrt2\,[\ln\mu + \ln(1-x)]\,.
 \ea
After using (\ref{xR}), and substracting a constant from the scalar
field, we obtain Eq. (\ref{phiG}), where the neglected terms are of
order $O(\mu^2) = O(\e^{-\hat{T}})$.

From (\ref{riscal}) the Ricci scalar is
 \be\lb{riscalA}
{\cal R} = -6l^{-2} - \frac2{b^2\sinh^4T} \simeq -
\frac1{8b^2(\mu-\nu)^2} \simeq - \frac{\e^{2\hat{T}}} {2(1-x)^4}\,,
 \ee
leading to
 \be\lb{r2riscal}
r^2{\cal R} = -2\alpha\frac{x^2}{(1-x)^2}\,.
 \ee

\newpage

\begin{figure}
\centering
\includegraphics[scale=0.7]{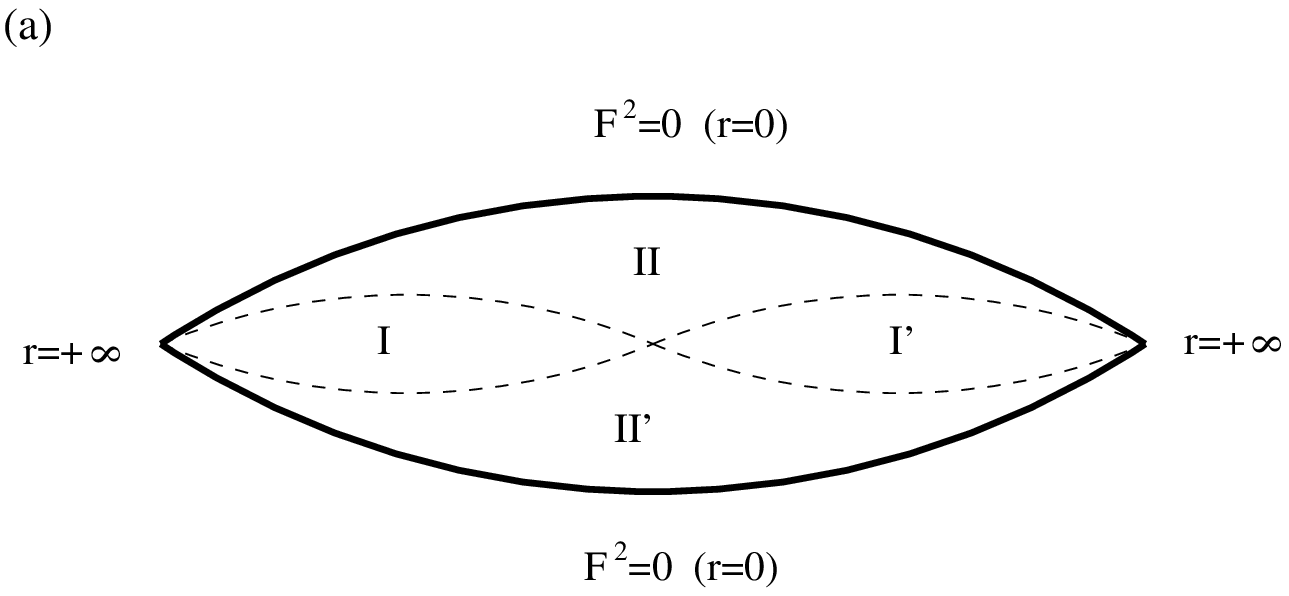}
\nonumber
\end{figure}

\begin{figure}
\centering
\includegraphics[scale=0.7]{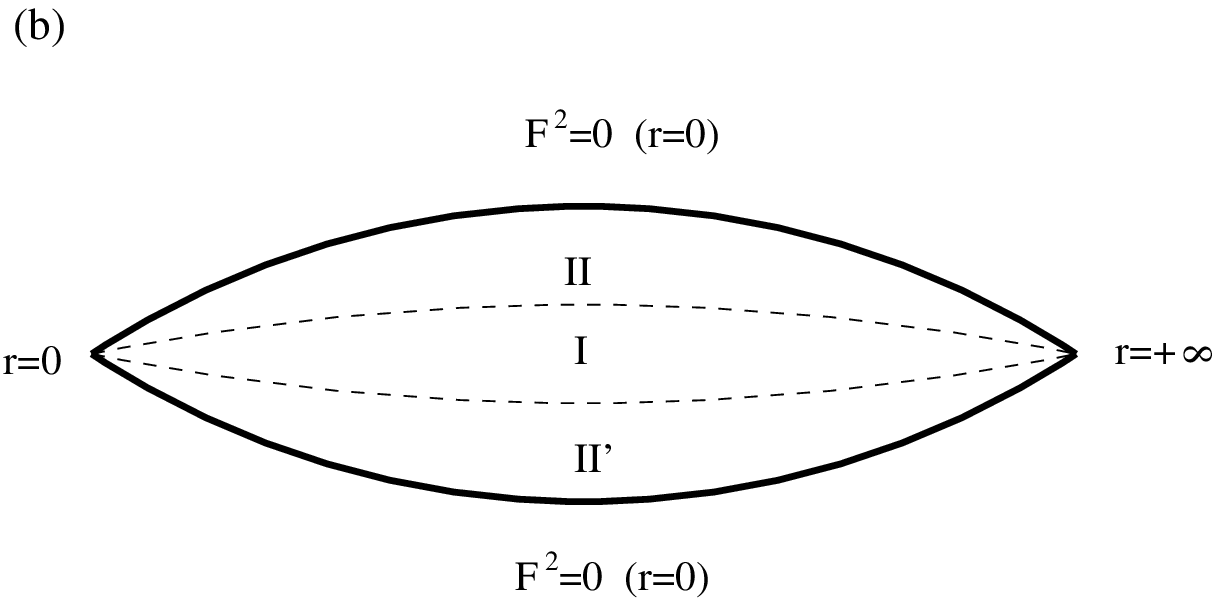}
\nonumber
\end{figure}

\begin{figure}
\centering
\includegraphics[scale=0.7]{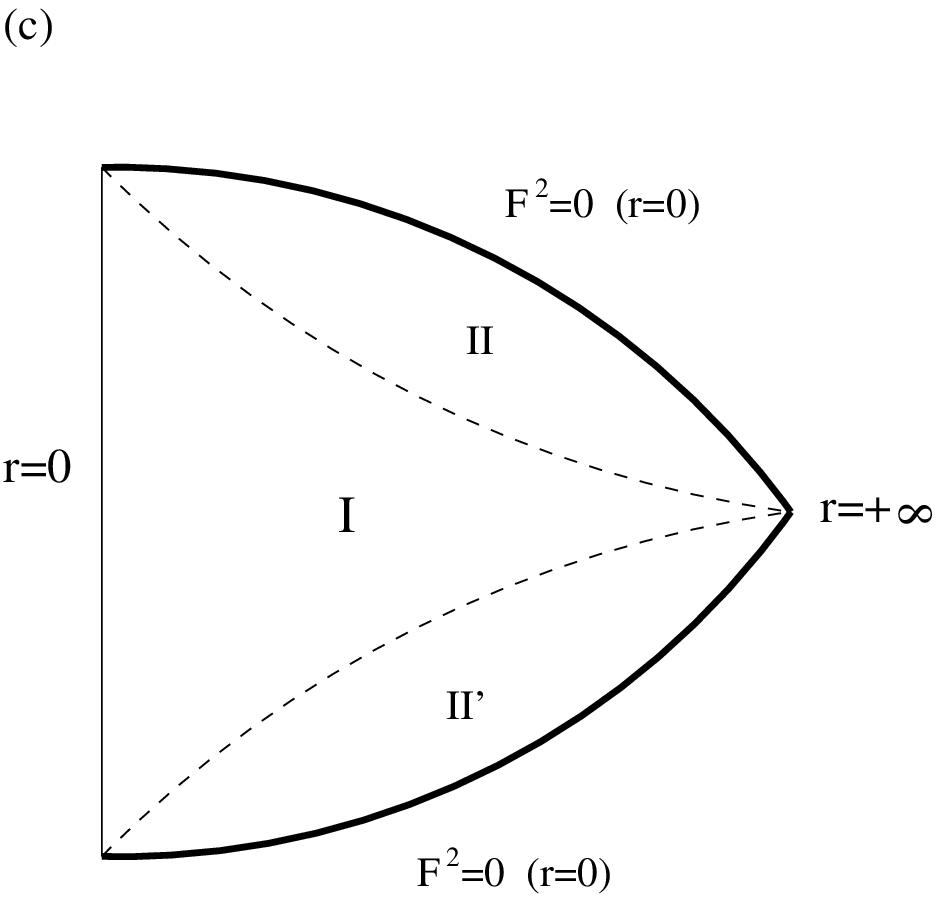}
\caption{Penrose diagram for the solution (\ref{etamet})} (a) for
$\alpha<0$, (b) for $\alpha=0$, (c) for $\alpha>0$. The curvature
singularity is shown as a heavy line, the apparent horizon as a
broken line.
\end{figure}

\end{document}